\documentclass{elsart}
\newcommand{\be}{\begin{equation}}
\newcommand{\ee}{\end{equation}}
\def\bea{\begin{eqnarray}}
\def\eea{\end{eqnarray}}
\def\gn{\gamma_\nu}
\def\g5{\gamma_5}
\def\vamn{\varepsilon^{a m n}}
\def\ot{(1\,\leftrightarrow\,2)}

\begin{document}

\begin{frontmatter}
\title
{ Lepton U$_l$(1) symmetry and interaction of neutrinos with deuterons at high 
energies\thanksref{talk}}
\thanks[talk]{Talk given at the 16th IUPAP International Conference on 
Few--Body Problems in Physics, March 6--10, 2000, Taipei, Taiwan}
\author[Prague]{\underline{E. Truhl\'{\i}k}} and  
\author[Edmonton]{F.C. Khanna}
\address[Prague]{Institute of Nuclear Physics, Academy of Sciences of the Czech Republic, CZ-25068
$\check{R}$e$\check{z}$, Czech Republic}
\address[Edmonton]{Theoretical Physics Institute, Department of Physics, University
of Alberta, Edmonton, Alberta,Canada,T6G 2J1 and
TRIUMF, 4004 Wesbrook Mall, Vancouver, BC, Canada, V6T 2A3 
}

\begin{abstract}
We discuss first the lepton U(1) symmetry and its relation to the presence of
the neutrino mass term in the Hamiltonian. 
Then we consider the possibility of the measurement of neutrino-deuteron reactions 
with high energy neutrinos at SNO by using neutrinos from a neutrino factory.
Next we report on a construction of weak axial one-boson exchange currents 
for the Bethe-Salpeter equation, starting from chiral Lagrangians. 
These Lagrangians are a natural extension of the non-linear $\sigma$ model
and they serve to describe processes at energies far from the threshold.
It is shown that the currents fulfil the Ward-Takahashi identities and the 
matrix element of the full current between the two-body solutions 
of the Bethe-Salpeter equation satisfies the PCAC constraint exactly.
Consistent calculations based on the proposed 
formalism would give more confidence in conclusions about the neutrino-deuteron
processes and consequently, about the neutrino oscillations. 
\end{abstract}
\end{frontmatter}
\section{Introduction}
\label{intro}
The physics of neutrino represents now a vast interdisciplinary area overlapping 
elementary particle-, nuclear-, astrophysics and cosmology. In the particle
classification, the neutrino belongs to leptons. There exists a consensus 
that its mass should not be large, maybe it is zero. 

With this in mind  and in an analogy with the other unitary symmetries \cite{TDL},
one is tempted to assume for the world of leptons the unitary U(1) symmetry
\be
U(1)\,=\,e^{i\,L\,\theta}\,,   \label{Ul}
\ee
where L is the lepton number. If the Hamiltonian H of the system is invariant under the
transformation
\be
U\,H\,U^+\,=\,H\,,    \label{UHU}
\ee
then 
\be
[L\,,\,H]\,=\,0\,,   \label{CLH}
\ee
L is conserved and the phase $\theta$ is unobservable.

A more detailed analysis of the weak interaction Hamiltonian reveals  \cite{DGH,KP} that 
the Dirac mass term can be written as
\be
{\mathcal L}_D\,=\,- m \bar \nu \nu\,=\,-m(\bar \nu_L \nu_R\,+\,\bar \nu_R \nu_L)\,,  \label{DMT}
\ee
where the left- and right-handed neutrinos are
\be
\nu_L\,=\,\frac{1}{2}(1\,-\,\g5)\,\nu\,,\qquad \nu_R\,=\,\frac{1}{2}(1\,
+\,\g5)\,\nu\,. \label{nuLR}
\ee
The mass term (\ref{DMT}) is clearly invariant under the transformation (\ref{Ul}).
However, the problem is that the right-handed neutrinos are not observed experimentally.

On the other hand, one can construct for a Majorana neutrino \cite{DGH,KP}
\be
\nu_M\,\equiv\,\nu_L\,+\,\nu^{\,c}_L\,, \quad\, \nu^{\,c}_L\,\equiv\,(\nu_L)^{\,c}\,,
\quad\, \nu^{\,c}\,=\,C\,{\bar \nu}^T\,, \label{nuM}
\ee
the mass term
\be
{\mathcal L}_M\,=\,- m \bar\nu_M \nu_M\,=\,-m(\bar{\nu^{\,c}}_L \nu_L\,+\,\bar\nu_L \nu^{\,c}_L)\,,  
\label{MMT}
\ee
from $\nu_L$ alone, but this term is not invariant under the phase transformation
(\ref{Ul}) implying the lepton number non-conservation. However, the direct experimental verification
of the Majorana nature of the neutrino is lacking: the neutrinoless double 
$\beta$-decay
has not yet been observed, only lower bounds for the life time of some nuclei
were established \cite{Bi}.

On the other hand, admitting that the lepton U(1) symmetry is not an exact one 
(in contrast to the electromagnetic U(1) symmetry), one is allowed to speculate on the 
presence of the Majorana neutrino mass terms in the Hamiltonian. Let us note that 
one is allowed to do it also in the Dirac neutrino case, because there does not
exist experimental proof that the neutrino mass is zero: only upper bound in
the direct neutrino mass search is known at present \cite{Z}, with a value of
about 2.5 eV for the electron neutrino mass. About the same upper limit for the Majorana
neutrino mass is obtained in the new experiment  \cite{Da} on the neutrinoless
double $\beta$--decay  of \nuc{116}{Cd}.

In its turn, the presence of the neutrino mass term in the Hamiltonian can give rise to the phenomenon 
of neutrino oscillations \cite{DGH,KP,B}. Actually, these oscillations are rather convincingly seen
in experiments with the solar and atmospheric neutrinos such as Homestake,
SuperKamiokande, GALLEX and SAGE \cite{B1}. In terrestrial experiments,
only the result of LSND \cite{LSND} is interpreted in favour of the oscillations. 

\section{Neutrino oscillations and SNO experiment}   
\label{CH1}

The clean experimental investigation of the neutrino oscillations is offered by exploiting
the ratio of charged (CC) and neutral (NC) cross sections for the neutrino-deuteron reactions, 
\bea
\nu_l\,+\,d\,&\longrightarrow&\,l^-\,+\,p\,+\,p\,,\quad (CC)        \label{CC}  \\
\nu_x\,+\,d\,&\longrightarrow&\,\nu_x\,+\,n\,+\,p\,,\quad (NC)  \label{NC}   \\
\nu_x\,+\,e^-\,&\longrightarrow&\,\nu_x\,+\,e^-\,,       \label{nue}
\eea
where $\nu_x$ refers to any active flavour of the neutrino.

These reactions are important
for studying the solar electron neutrino oscillations (l\,=\,e) and they are the main object
of the SNO detector \cite{SNO} at present. The aim of the detector is to compare the flux of the 
electron neutrinos produced in the sun presumably by the reaction
\be
^8B\,\longrightarrow\,^8Be^*\,+\,e^{\,+}\,+\,\nu_e\,,\qquad E_\nu\,\le\,15\,MeV\,, \label{8Bnu}
\ee
to the flux of all active flavours of the 
neutrinos. Deviation of this ratio from the prediction based on the Standard Model
with the ansatz of zero neutrino mass would provide a strong support for the massive neutrinos,
but not for the violation of the lepton U(1) symmetry. As mentioned above, the fate of this symmetry
is strongly correlated with the experimental evidence of the nuclear neutrinoless double 
$\beta$-decay.

Among other experimental activities, SNO intends to measure the flavour composition of the 
atmospheric neutrino flux \cite{SNO}. The energies of these neutrinos can be much larger
than for the solar neutrinos (up to 10 GeV and more).

Another investigative potential of SNO is related to the nowadays discussed neutrino factory
\cite{G,C,McD}. The idea is to create muon storage rings which would produce the neutrino
beam by allowing muons to decay in the stright section of a storage ring. The obtained beam
woud have a precisely known composition depending only on the muons which decay, 
\bea
\mu^{\,+}\,&\longrightarrow&\,e^{\,+}\,+\,\nu_e\,+\,\bar \nu_\mu\,,  \label{mup}  \\
\mu^{\,-}\,&\longrightarrow&\,e^{\,-}\,+\,\bar \nu_e\,+\,\nu_\mu\,.  \label{mum} 
\eea
In comparison with the low-energy solar neutrinos, the neutrinos from reactions (\ref{mup}),
(\ref{mum}) can be produced at much higher energy interval (up to 50 GeV at present),
thus giving another powerful method to investigate the transition probability 
$\nu_e\,(\bar \nu_e)\,\rightarrow\,\nu_\mu\,(\bar \nu_\mu)$ by searching for leptons
with the opposite sign ("wrong-sign" leptons). Besides reactions 
(\ref{CC})-(\ref{nue}), we now also have similar reaction for the antineutrinos,
\bea
\bar \nu_l\,+\,d\,&\longrightarrow&\,l^+\,+\,n\,+\,n\,,        \label{bCC}  \\
\bar \nu_x\,+\,d\,&\longrightarrow&\,\bar \nu_x\,+\,n\,+\,p\,,  \label{bNC}   \\
\bar \nu_x\,+\,e^-\,&\longrightarrow&\,\bar \nu_x\,+\,e^-\,.       \label{bnue}
\eea

The estimated neutrino flux  is $\sim$ \mbox{10$^{\,10}$\,-\,10$^{\,11}$ $\bar \nu_\mu$
m$^{-2}$ year$^{-1}$} \cite{G} from the positive muons stored with 
the momenta p\,=\,20\,-\,50 GeV/c. In such a factory, $\nu_e$
 are produced with about the same flux  \cite{G}.
It is to be noted that the flux of the atmospheric neutrinos at the terrestrial surface is 
$\sim$\,3\,x\,10$^{10}$ m$^{-2}$ year$^{-1}$ \cite{A}, which is close to the estimated flux
of the neutrinos from the neutrino factory.

These considerations make theoretical studies of high energy neutrinos
with the deuterons interesting. The cross sections for the neutrino-deuteron reactions
up to energies 170 MeV have been already calculated \cite{KN} within the framework of the standard
nuclear physics calculations at low energies, however, with the static part of the
weak exchange current of the pion range included.

\section{Meson exchange currents for the Bethe-Salpeter equation}
\label{CH2}

At high energies, relativistic effects both in the potential kernel and transition 
operator (current) should be considered within the same formalism. 
This can be done from the beginning within the framework of relativistic equations. 
As a step in this direction, we have recently carried out \cite{KT1} an investigation of 
the structure of the weak axial meson exchange currents (MEC)
in conjunction with the Bethe-Salpeter (BS) equation considered in the ladder approximation.
Both the OBE potentials $\hat V_B$ and the one- and two-nucleon current operators were constructed
from chiral Lagrangians of the N$\Delta(1236)\pi \rho\,$a$_1 \omega$ system and from the
associated one-body currents using the technique of Feynman diagrams. These Lagrangians
are an extension of the standard non-linear $\sigma$ model \cite{DGH}, which is invariant
under the transformations from the global chiral group SU(2)$_L\times $SU(2)$_R$. This model
contains nucleons and pions only.

One of the employed Lagrangians \cite{IT} represents the appraoch developed earlier, 
in which the heavy meson fields $\rho$ and a$_1$ are introduced as massless 
Yang-Mills (YM) compensating fields. 
These fields belong to the linear realization of the local chiral SU(2)$_L\times $SU(2)$_R$ 
symmetry. Actually, this symmetry is violated by the heavy meson mass terms introduced 
by hands. Also the external electroweak interactions should be introduced by hand, 
since there are no other charges available, which could be associated with them. 
Inspite of this internal inconsistency, this approach turned out to be succesful in 
describing the nuclear phenomena in the region of low and intermediate energies \cite{EW}.

The defects of the approach of the massive YM fields are removed in the approach of 
hidden local symmetries (HLS) \cite{BKY,M}. 
In this method, a given global symmetry group G$_g$
of a system Lagrangian is extended to a larger one by a local group
H$_l$ and the Higgs mechanism generates the mass terms for gauge fields
of the local group in such a way that the local symmetry is preserved.
For the chiral group G$_g\,\equiv$\,[SU(2)$_L \times $SU(2)$_R$]$_g$ and
H$_l\, \equiv$\,[SU(2)$_L \times $SU(2)$_R$]$_l$ the gauge particles are
identified \cite{BKY,M} with the $\rho$- and a$_1$ mesons.
An additional extension by a local group U$_l$(1) allows one to include
the isoscalar $\omega$ meson as well \cite{KM}. Moreover, external
gauge fields, which are related  to the electroweak interactions
of the Standard Model, are included by gauging the
global chiral symmetry group G$_g$. Lagrangian constructed within the HLS scheme
and suitable for constructing the exchange currents is given in \cite{STG}.

The physical content of our YM and HLS Lagrangians differs only due to the different choice
of higher order terms in the Lagrangian correcting the high energy behaviour of elementary
amplitudes.

The Lagrangians contain all necessary vertices which together with the associated
one-body currents can be used in constructing many body current operators.
We restrict ourselves to the one- and two-nucleon weak axial MEC.

In our approach, the operator of the one-nucleon weak axial current for 
the i{\it th} nucleon is

\be
\hat{J}^{a\,\mu}(1,i)\,=\,\frac{g_A}{2}\,m^2_{a_1}\,\Delta^{\mu\,\nu}_{a_1}(q)\,
\left( \gn \g5 \tau^a \right )_i\,
-\,gf_\pi\,\Delta^\pi_F(q^2)\,q^\mu\,\left (\g5 \tau^a \right )_i\,. \label{ONAC}
\ee
Here g$_A$\,=\,--\,1.26 and  $\Delta^{\mu\,\nu}_{a_1}(q)$ ($\Delta^\pi_F(q^2)$)
is the propagator of the a$_1$ ($\pi$) meson.

The divergence of the current (\ref{ONAC}) is
\be
q_\mu \hat{J}^{a\,\mu}(1,i)\,=\,\left[\hat{e}_A(i),\,G^{-1}_i\right]_+\,
-\,\,gf_\pi m^2_\pi\,\Delta^\pi_F(q^2)\,\left (\g5 \tau^a \right )_i
                  \,. \label{DOBCi}
\ee
Here  the operator of the nucleon axial charge $\hat{e}_A$(i) and the inverse of the
nucleon propagator G$^{-1}$(p) are defined as
\be
\hat{e}_A(i)\,=\,g_A\,\left(\g5\frac{\tau^a}{2}\right)_i\,,\quad G^{-1}(\,p\,)\,=\,
\not p\,-\,M\,. \label{def1}
\ee

Generally, a MEC constructed in the conjuction with the BS  contains various contact 
and mesonic terms, but it does not contain the nucleon Born terms, the contribution from 
which is actually generated when one calculates the matrix element of the one-nucleon 
current between the solutions of the BS equation. The general structure of the weak axial 
MEC operator is given in Fig.~1.

The details of the structure of the constructed  weak axial MEC operators can be 
found in \cite{KT1}. Here we discuss the essential features of their  structure.

a) The currents derived from the YM type Lagrangian

For the B=$\pi$, $\omega$ meson exchanges we have
\begin{equation}
q_\mu \hat{J}^{a\,\mu}_{BS\,B}(ex)\,=\,\left[\,\hat{e}_A(1)+\hat{e}_A(2),\,\hat{V}_B\,\right]_+\,
+\,if_\pi\,m^2_\pi\Delta^\pi_F(q^2)\,\hat{\mathcal M}^a_B(2)\,,  \label{divex}
\end{equation}
where $\hat{\mathcal M}^a_B(2)$ is the two-body pion absorption amplitude.

In this model, the $\rho$ and $a_1$ meson exchanges should be considered together and the 
resulting MEC operator satisfies the continuity equation
\begin{equation}
q_\mu \hat{J}^{a\,\mu}_{BS\,\rho + a_1}(ex)\,=\,
 \left[\hat{e}_A(1)\,+\,\hat{e}_A(2)\,,\,\hat{V}_\rho\,+\,\hat{V}_{a_1} \right]_+\, 
+\,i f_\pi m^2_\pi\,\Delta^\pi_F(q^2)\,\hat {\mathcal M}^a_{\rho + a_1}(2)\,.   \label{dJRPA1EXBS}
\end{equation}

The obtained MECs differ from those constructed for the on--shell nucleons by the presence
of some additional terms which disappear for the on--shell nucleons. 
As an example, let us consider 
the contribution from one of the pion contact terms,
\bea
\hat{J}^{a\,\mu}_{c_1\,\pi}(\pi)\,&\equiv&\,if_\pi\,q^\mu
\Delta^\pi_F(q^2)\,\hat{\mathcal M}^a_{c_1\,\pi}\,, \label{Jppp} \\
\Delta\hat{J}^{a\,\mu}_{c_1\,\pi}(\pi)\,&\equiv&\,if_\pi\,q^\mu
\Delta^\pi_F(q^2)\,\Delta \hat{\mathcal M}^a_{c_1\,\pi}\,, \label{dJppp}
\eea
where the pion absorption amplitudes are
\bea
\hat{\mathcal M}^a_{c_1\,\pi}\,&=&\,-\frac{i}{2}\left( \frac{g}{M} \right)^2\,
\vamn \not q_2 \tau^m_1 \Delta^\pi_F(q^2_2)\,\hat{\Gamma}^n_2\,+\,\ot\,, \label{Mappp} \\
\Delta \hat{\mathcal M}^a_{c_1\,\pi}\,&=&\,-\frac{i}{4}\left( \frac{g}{M} \right)^2\,
\vamn \not q_1 \tau^m_1 \Delta^\pi_F(q^2_2)\,\hat{\Gamma}^n_2\,+\,\ot\,, \label{dMappp} 
\eea
Besides the contact current (\ref{Jppp}), constructed earlier in \cite{TK2}
we have an additional term (\ref{dJppp}) which disappears for the on-shell  nucleons.
This can be seen from (\ref{dMappp}), where the right hand side is proportional to
the momentum transfer $\not q_1$ which provides zero when sandwiched between the
spinors of the 1{\it st} nucleon.

b) For the currents derived from the HLS type Lagrangian, Eq.~(\ref{divex}) holds
for all considered exchanges.

The full BS weak axial current is defined as 
\be
\hat{J}^{a\,\mu}_{BS}\,=\,\hat{J}^{a\,\mu}_{IA}\,+\,\hat{J}^{a\,\mu}_{BS}(ex)\,, \label{JBSEH}
\ee
where the impulse approximation current is
\be
\hat{J}^{a\,\mu}_{IA}\,=\,i\hat{J}^{\mu}_a(1,1)G^{-1}_2\,+\,i\hat{J}^{\mu}_a(1,2)G^{-1}_1
\,, \label{JIA}
\ee
and the weak axial MEC $\hat{J}^{a\,\mu}_{BS}(ex)$ is given by the contributions
from the  exchanges considered in \cite{KT1}.

Using the Ward-Takahashi identities  for the one- and two-nucleon currents \cite{KT1} 
yields for the divergence of the full BS current
\begin{equation}
q_\mu \hat{J}^{a\,\mu}_{BS}\,=\,[\,\hat{e}_A(1)+\hat{e}_A(2),\,{\mathcal G}^{-1}\,]_+\,
+\,if_\pi\,m^2_\pi\Delta^\pi_F(q^2)\,\hat{\mathcal M}^a\,,  \label{divf}
\end{equation}
where the inverse Green function is 
\be
{\mathcal G}^{-1}\,=\,G^{-1}_{BS}\,+\,\hat V\,,  \label{IGF}
\ee 
$\hat{\mathcal M}^a$ is the full pion absorption amplitude
\be
\hat{\mathcal M}^a\,=\,i\hat{\Gamma}^a_1 G^{-1}_2+i\hat{\Gamma}^a_2 G^{-1}_1+
\hat{\mathcal M}^a(2)\,,\quad \hat{\Gamma}^a_i\,=\,ig\left (\g5 \tau^a \right )_i\,   \label{Ma}
\ee
the BS propagator in term of the single-particle propagators reads $G_{BS}=-iG_1 G_2$
and the full potential is, 
\be
\hat V\,=\,\hat V_\pi\,+\,\hat V_\rho\,+\,\hat V_{a_1}\,+\,\hat V_\omega\,. \label{V} 
\ee
Because the two-body BS wave functions for both bound and scattering states satisfy
the equation
\be
{\mathcal G}^{-1}|\psi>\,=\,<\psi|{\mathcal G}^{-1}\,=\,0\,,  \label{BSE}
\ee
the matrix element of the divergence of the full current (\ref{divf}) fulfil
the standard PCAC constraint
\begin{equation}
q_\mu <\psi|\,\hat{J}^{a\,\mu}_{BS}\,|\psi>\,=\,if_\pi\,m^2_\pi\Delta^\pi_F(q^2)\,
<\psi|\,\hat{\mathcal M}^a\,|\psi>\,.  \label{divme}
\end{equation}

The model dependence is given by effects coming from the difference between the 
MECs derived in considered schemes. More detailed checking \cite{KT1} shows that
they are either of short-range nature or momentum dependent. 

\section{Results and conclusions}
\label{RC}

We constructed the weak axial MECs of the $\pi$, $\rho$, a$_1$ and $\omega$ range 
ready to use with the BS equation. Two different chiral schemes were chosen as a starting 
point to get a realistic set of currents. The full BS currents satisfy the Ward-Takahashi 
identity and the divergence of their matrix elements between the two-body BS wave
functions satisfies the standard PCAC constraint.

The constructed currents can be immediately used to improve the recent calculations 
\cite{KKG} of the solar proton burning process,
\be
p\,+\,p\,\longrightarrow\,d\,+\,e^{\,+}\,+\,\nu_e\,,   \label{ppd}
\ee
within the BS formalism. As the next step would be consistent covariant calculations 
of the cross sections for the neutrino-deuteron reactions (\ref{CC}),(\ref{NC}) and 
(\ref{bCC}),(\ref{bNC}) 
with energetic neutrinos coming from the atmosphere or from a neutrino factory. 
The consistent calculations would provide a possibility to compare the 
covariant BS- and standard nuclear physics calculations, thus bringing them under control. 
Such calculations would give more confidence in conclusions about the neutrino oscillations
being investigated at SNO.



\section*{Acknowledgments}
The work of E.~T.~is supported by the grant GA \v{C}R 202/00/1669 and by the
grant of NSC ROC.
Research of F.~C.~K.~is supported in part by NSERCC.


\input feynman
\newpage
\hspace{130pt}
\begin{picture}(10000,15000)
\drawline\fermion[\N\REG](0,0)[13948]
\drawarrow[\N\ATBASE](\pmidx,\pmidy)
\drawarrow[\N\ATBASE](0,2000)
\drawarrow[\N\ATBASE](0,11948)
\put(2450,-1800){a}
\global\advance\pmidx by 450
\global\advance\pmidy by -300
\put(\pmidx,\pmidy){$P$}
\global\advance\pmidx by -2000
\put(\pmidx,\pmidy){$\mathcal N$}
\global\advance\fermionbackx by -1500
\put(\fermionbackx,\fermionbacky){$p_{\,1}^{\,\prime}$}
\put(\fermionbackx,0){$p_{\,1}$}
\global\advance\pmidy by -2974
\thicklines
\drawline\fermion[\W\REG](0,\pmidy)[2000]
\drawarrow[\E\ATBASE](\pmidx,\pmidy)
\global\advance\pmidx by -600
\global\advance\pmidy by 400
\put(\pmidx,\pmidy){$B$}
\thinlines
\drawline\photon[\W\REG](\fermionbackx,\fermionbacky)[4]
\global\advance\pmidx by -1600
\global\advance\pmidy by 800
\put(\pmidx,\pmidy){$\hat{J}^{a,\,\mu}(q)$}
\thicklines
\global\seglength=1400
\global\gaplength=350
\drawline\scalar[\E\REG](0,9948)[3]
\drawarrow[\E\ATBASE](\pmidx,\pmidy)
\global\advance\pmidx by -450
\thinlines
\global\advance\pmidy by 400
\put(\pmidx,\pmidy){$B_2$}
\global\advance\pmidy by -1300
\put(\pmidx,\pmidy){$q_{\,2}$}
\drawline\fermion[\N\REG](4900,0)[13948]
\drawarrow[\N\ATBASE](4900,2000)
\drawarrow[\N\ATBASE](4900,11948)
\global\advance\fermionbackx by 600
\put(\fermionbackx,\fermionbacky){$p_{\,2}^{\,\prime}$}
\put(\fermionbackx,0){$p_{\,2}$}
\end{picture}
\hspace{20pt}
\begin{picture}(10000,15000)
\drawline\fermion[\N\REG](0,0)[13948]
\drawarrow[\N\ATBASE](\pmidx,\pmidy)
\drawarrow[\N\ATBASE](0,2000)
\drawarrow[\N\ATBASE](0,11948)
\put(2450,-1800){b}
\global\advance\pmidx by 450
\global\advance\pmidy by -300
\put(\pmidx,\pmidy){$Q$}
\global\advance\pmidx by -2000
\put(\pmidx,\pmidy){$\mathcal N$}
\global\advance\fermionbackx by -1500
\put(\fermionbackx,\fermionbacky){$p_{\,1}^{\,\prime}$}
\put(\fermionbackx,0){$p_{\,1}$}
\global\advance\pmidy by 2974
\thicklines
\drawline\fermion[\W\REG](0,\pmidy)[2000]
\drawarrow[\E\ATBASE](\pmidx,\pmidy)
\global\advance\pmidx by -600
\global\advance\pmidy by 400
\put(\pmidx,\pmidy){$B$}
\thinlines
\drawline\photon[\W\REG](\fermionbackx,\fermionbacky)[4]
\global\advance\pmidx by -1600
\global\advance\pmidy by 800
\put(\pmidx,\pmidy){$\hat{J}^{a,\,\mu}(q)$}
\thicklines
\global\seglength=1400
\global\gaplength=350
\drawline\scalar[\E\REG](0,2974)[3]
\drawarrow[\E\ATBASE](\pmidx,\pmidy)
\global\advance\pmidx by -450
\thinlines
\global\advance\pmidy by 400
\put(\pmidx,\pmidy){$B_2$}
\global\advance\pmidy by -1300
\put(\pmidx,\pmidy){$q_{\,2}$}
\drawline\fermion[\N\REG](4900,0)[13948]
\drawarrow[\N\ATBASE](4900,2000)
\drawarrow[\N\ATBASE](4900,11948)
\global\advance\fermionbackx by 600
\put(\fermionbackx,\fermionbacky){$p_{\,2}^{\,\prime}$}
\put(\fermionbackx,0){$p_{\,2}$}
\end{picture}
\vspace{50pt}
\newline
\hspace{50pt}
\begin{picture}(10000,15000)
\drawline\fermion[\N\REG](0,0)[13948]
\drawarrow[\N\ATBASE](0,2000)
\drawarrow[\N\ATBASE](0,11948)
\put(2450,-1800){c}
\global\advance\fermionbackx by -1500
\put(\fermionbackx,\fermionbacky){$p_{\,2}^{\,\prime}$}
\put(\fermionbackx,0){$p_{\,2}$}
\global\advance\fermionbackx by 1500
\thicklines
\global\seglength=1400
\global\gaplength=350
\drawline\scalar[\E\REG](0,6974)[3]
\thinlines
\drawarrow[\W\ATBASE](\pmidx,\pmidy)
\global\advance\pmidx by -450
\global\advance\pmidy by 400
\put(\pmidx,\pmidy){$B_2$}
\global\advance\pmidy by -1300
\put(\pmidx,\pmidy){$q_{\,2}$}
\drawline\fermion[\N\REG](4900,0)[13948]
\drawarrow[\N\ATBASE](4900,2000)
\drawarrow[\N\ATBASE](4900,11948)
\global\advance\fermionbackx by 600
\put(\fermionbackx,\fermionbacky){$p_{\,1}^{\,\prime}$}
\put(\fermionbackx,0){$p_{\,1}$}
\thicklines
\drawline\fermion[\E\REG](\pmidx,\pmidy)[2000]
\drawarrow[\W\ATBASE](\pmidx,\pmidy)
\global\advance\pmidx by -600
\global\advance\pmidy by 400
\put(\pmidx,\pmidy){$B$}
\thinlines
\drawline\photon[\E\REG](\fermionbackx,\fermionbacky)[4]
\global\advance\pmidx by -1600
\global\advance\pmidy by 800
\put(\pmidx,\pmidy){$\hat{J}^{a,\,\mu}(q)$}
\end{picture}
\hspace{50pt}
\begin{picture}(10000,15000)
\drawline\fermion[\N\REG](0,0)[13948]
\drawarrow[\N\ATBASE](0,2000)
\drawarrow[\N\ATBASE](0,11948)
\put(4200,-1800){d}
\global\advance\fermionbackx by -1500
\put(\fermionbackx,\fermionbacky){$p_{\,1}^{\,\prime}$}
\put(\fermionbackx,0){$p_{\,1}$}
\thicklines
\global\seglength=1400
\global\gaplength=350
\drawline\scalar[\E\REG](0,9948)[5]
\thinlines
\drawarrow[\E\ATBASE](6000,\pmidy)
\drawarrow[\W\ATBASE](2400,\pmidy)
\put(1800,10450){$B_1$}
\put(6300,10450){$B_2$}
\put(1800,9000){$q_{\,1}$}
\put(6300,9000){$q_{\,2}$}
\drawline\photon[\S\REG](\pmidx,\pmidy)[7]
\global\advance\pmidx by 200
\global\advance\pmidy by -1100
\drawarrow[\N\ATBASE](\pmidx,\pmidy)
\global\advance\pmidx by -1600
\global\advance\pmidy by -3300
\put(\pmidx,\pmidy){$\hat{J}^{a,\,\mu}(q)$}
\drawline\fermion[\N\REG](8400,0)[13948]
\drawarrow[\N\ATBASE](8400,2000)
\drawarrow[\N\ATBASE](8400,11948)
\global\advance\fermionbackx by 600
\put(\fermionbackx,\fermionbacky){$p_{\,2}^{\,\prime}$}
\put(\fermionbackx,0){$p_{\,2}$}
\end{picture}
\hspace{50pt}
\begin{picture}(10000,15000)
\drawline\fermion[\N\REG](0,0)[13948]
\drawarrow[\N\ATBASE](0,2000)
\drawarrow[\N\ATBASE](0,11948)
\global\advance\fermionbackx by -1500
\put(\fermionbackx,\fermionbacky){$p_{\,1}^{\,\prime}$}
\put(\fermionbackx,0){$p_{\,1}$}
\put(4200,-1800){e}
\global\seglength=1400
\global\gaplength=350
\thicklines
\drawline\scalar[\E\REG](0,9948)[5]
\thinlines
\drawarrow[\E\ATBASE](6000,\pmidy)
\drawarrow[\W\ATBASE](2400,\pmidy)
\put(1800,10450){$B_1$}
\put(6300,10450){$B_2$}
\put(1800,9000){$q_{\,1}$}
\put(6300,9000){$q_{\,2}$}
\thicklines
\drawline\fermion[\S\REG](\pmidx,\pmidy)[2100]
\drawarrow[\N\ATBASE](\pmidx,\pmidy)
\global\advance\pmidx by 300
\global\advance\pmidy by -400
\put(\pmidx,\pmidy){$B$}
\thinlines
\drawline\photon[\S\REG](\fermionbackx,\fermionbacky)[5]
\global\advance\pmidx by -1600
\global\advance\pmidy by -3300
\put(\pmidx,\pmidy){$\hat{J}^{a,\,\mu}(q)$}
\drawline\fermion[\N\REG](8400,0)[13948]
\drawarrow[\N\ATBASE](8400,2000)
\drawarrow[\N\ATBASE](8400,11948)
\global\advance\fermionbackx by 600
\put(\fermionbackx,\fermionbacky){$p_{\,2}^{\,\prime}$}
\put(\fermionbackx,0){$p_{\,2}$}
\end{picture}

\vspace{10mm}
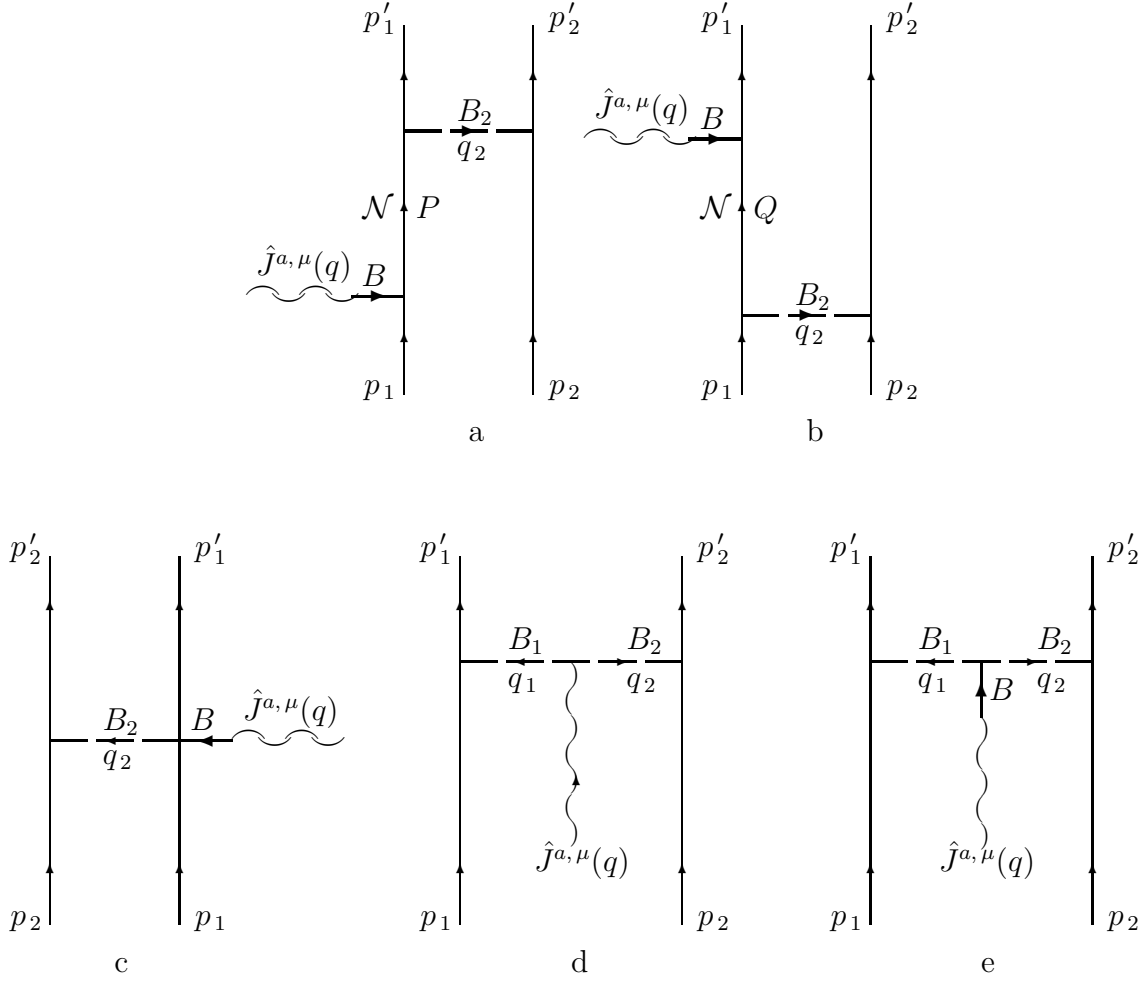
Fig.\,1. The general structure of the weak axial MEC operators considered.
The weak axial
interaction is mediated by the meson B which is either $\pi$ or $a_1$ meson.
The range of the
current is given by the meson $B_2$.
The graphs a, b, represent the current ${\hat J}^{a\,\mu}_{B_2}({\mathcal N},\,B)$ with
$\mathcal N$
either for the nucleon N (nucleon Born terms)  or for the $\Delta(1236)$ isobar.
In conjunction with the BS equation, the nucleon Born terms do not enter the
MEC.
The graph c represents a contact current ${\hat J}^{a\,\mu}_{c\,B_2}(B)$.
Another type of the contact
terms is given by the graph d, ${\hat J}^{a\,\mu}_{B_1\,B_2}$, where the weak axial
current interacts directly
with the mesons $B_1$ and $B_2$.
The graph e is for a mesonic current ${\hat J}^{a\,\mu}_{B_1\,B_2}(B)$.
The associated pion absorption amplitudes correspond to the graphs where
the weak axial interaction
is mediated by the pion, but with the weak interaction wavy line removed.

\end{document}